\documentclass[prd,showpacs,preprintnumbers,amsmath,amssymb]{revtex4}
\usepackage{graphicx}% Include figure files
\usepackage{dcolumn}% Align table columns on decimal point
\usepackage{bm}% bold math
\usepackage{psfig}
\usepackage{epsfig}

\begin{document}

\newcommand{\vAi}{{\cal A}_{i_1\cdots i_n}}
\newcommand{\vAim}{{\cal A}_{i_1\cdots i_{n-1}}}
\newcommand{\vAbi}{\bar{\cal A}^{i_1\cdots i_n}}
\newcommand{\vAbim}{\bar{\cal A}^{i_1\cdots i_{n-1}}}
\newcommand{\htS}{\hat{S}}
\newcommand{\htR}{\hat{R}}
\newcommand{\htI}{\hat{I}}
\newcommand{\htB}{\hat{B}}
\newcommand{\htD}{\hat{D}}
\newcommand{\htV}{\hat{V}}
\newcommand{\cT}{{\cal T}}
\newcommand{\cM}{{\cal M}}
\newcommand{\cMs}{{\cal M}^*}
\newcommand{\vk}{{\bf k}}
\newcommand{\vK}{{\\bf K}}
\newcommand{\vb}{{\textstyle{\bf b}}}
\newcommand{{\vp}}{{\vec p}}
\newcommand{{\vq}}{{\vec q}}
\newcommand{\vQ}{{\vec Q}}
\newcommand{\vx}{{\textstyle{\bf x}}}
\newcommand{\tr}{{{\rm Tr}}}
\newcommand{\beq}{\begin{equation}}
\newcommand{\eeq}[1]{\label{#1} \end{equation}}
\newcommand{\half}{{\textstyle \frac{1}{2} }}
\newcommand{\lton}{\mathrel{\lower.9ex \hbox{$\stackrel{\displaystyle
<}{\sim}$}}}
\newcommand{\gton}{\mathrel{\lower.9ex \hbox{$\stackrel{\displaystyle
>}{\sim}$}}}
\newcommand{\ee}{\end{equation}}
\newcommand{\ben}{\begin{enumerate}}
\newcommand{\een}{\end{enumerate}}
\newcommand{\bit}{\begin{itemize}}
\newcommand{\eit}{\end{itemize}}
\newcommand{\bc}{\begin{center}}
\newcommand{\ec}{\end{center}}
\newcommand{\bea}{\begin{eqnarray}}
\newcommand{\eea}{\end{eqnarray}}
\newcommand{\beqar}{\begin{eqnarray}}
\newcommand{\eeqar}[1]{\label{#1}\end{eqnarray}}
\newcommand{\bra}[1]{\langle {#1}|}
\newcommand{\ket}[1]{|{#1}\rangle}
\newcommand{\norm}[2]{\langle{#1}|{#2}\rangle}
\newcommand{\brac}[3]{\langle{#1}|{#2}|{#3}\rangle}
\newcommand{\hilb}{{\cal H}}
\newcommand{\pleft}{\stackrel{\leftarrow}{\partial}}
\newcommand{\pright}{\stackrel{\rightarrow}{\partial}}

\begin{flushright}
%DRAFT 4/7 10pm
%\vskip .5cm
\end{flushright} \vspace{1cm}

\title{Consistent Determination of Particle Production in p-A, d-A and A-A Collisions using Color Glass Condensate}

\author{A.~Adil}%
\email{azfar@phys.columbia.edu}

\author{M.~Gyulassy}
\email{gyulassy@fias.uni-frankfurt.de}
%\email{gyulassy@phys.columbia.edu}

%\author{T.~Hirano}
%\email{hirano@phys.columbia.edu}

\affiliation{Frankfurt Institute for Advanced Studies (FIAS), 60438 Frankfurt am Main, Germany}
\affiliation{Department of Physics, Columbia University,
             538 West 120$^{th}$ Street, New York, NY 10027, USA}
%\affiliation{Columbia University, Department of
%Physics, 538 West 120-th Street, New York, NY 10027
%}%

\date{\today}% It is always \today, today,

\begin{abstract}
The success of nonviscous hydrodynamics in describing
the collective flow properties of bulk low $p_\perp$ 
observables at RHIC has led to the claim that a novel form of {\it strongly
coupled} Quark Gluon Plasma (sQGP) is created in 200 AGeV Au+Au
collisions.  This success  depends strongly, however, on the initial
conditions assumed in the calculations.  In particular, agreement with
data is only obtained assuming Glauber nuclear reaction plane (participant) geometry.
The KLN model of Color Glass Condenstate (CGC) initial conditions require
the existence of nonvanishing viscous effects.
We develop an improved model of  CGC that is more internally
consistent between the central and diffuse edge regions.
The improved model is shown to 
describe bulk rapidity distributions
for a wide assortment of system
sizes, geometries and energies.  The self consistency forces a specific $\sqrt{s}$ dependence of the multiplicity which leads to surprisingly low particle production predictions for Pb+Pb collisions at LHC COM energies.  These predictions are similar to those made using simple linear extrapolations of multiplicity attempted by the PHOBOS collaboration.
%at a correct description of smaller
%systems is likely to require a fully fluctuating monte carlo model for
%particle production.
 \end{abstract}


\maketitle

%%%%%%%%%%%%%%%%%%%%%%%%%%%%%%%%%%%%%%%%%%%%%%%%%%%%%%%%%%%%%%%%%%%%%%%%%%
\section{Introduction}

One of the most exciting discoveries in high energy nucleus-nucleus
collisions at the Relativistic Heavy Ion Collider (RHIC) is the large
magnitude of the elliptic flow parameter $v_2(p_\perp)$.  $v_2=\langle \cos(2\phi)\rangle$ characterizes the transverse
azimuthal distribution $dN/dp_\perp d\phi$.  The impact parameter
or participant nucleon dependence of 
$\langle v_2 \rangle$ provides a clear 
signature of the strong collective behavior of the created
medium~\cite{Sorensen:2003kp,Adams:2003am,Adler:2003kt,Adcox:2004mh}.
The systematics of $v_2$ (with energy, nuclear size, and centrality)
measured  at RHIC is
consistent with a description of the created Quark Gluon Plasma (QGP)
core (with $T(x)>T_c\approx 170$ MeV) as a non-viscous ``perfect'' 
fluid that evolves under Euler hydrodynamic equations of
motion  from an early thermalization 
time $\tau\sim 0.6$ fm ~\cite{Huovinen:2001cy,Heinz:2004ar} 
until the hadronization hypersurface
($T(x)=T_c$) is reached.
Below $T_c$ the highly viscous hadronic resonance gas ``corona''
evolves with conventional hadronic transport
~\cite{Hirano:2004rs,Hirano:2005xf}.  Such a hybrid perfect fluid sQGP core
plus nonequilibrium hadron resonance gas corona description,
however, is particularly 
sensitive to the assumed initial geometric distribution
of partons produced in nuclear collisions since
$v_2\propto \epsilon_\perp=\langle x^2-y^2 \rangle/\langle x^2+y^2 \rangle$.  Here, the averaging $\langle \rangle $ is over the transverse plane where $x$ and $y$ are the transverse spatial coordinates and the initial number density is used as the weight.
The success of the hybrid approach to describe the $v_2$ data
is contingent on the validity of the widely assumed Glauber model of
participant nucleon
geometry.

With Glauber initial conditions, the parton interaction cross
sections needed to generate the collective behavior seen at RHIC 
 are much larger than can be
supported via purely perturbative collisional
processes~\cite{Molnar:2001ux}.  This has led to the recent 
{\it strongly coupled} (s)QGP paradigm, which is supported
by the success of gauge-string dual Anti-deSitter/Conformal Field
Theory (AdS/CFT)~\cite{Maldacena:1997re} models in describing the low
shear viscosity implied by RHIC
data~\cite{Policastro:2001yc,Policastro:2002se,Policastro:2002tn}.

Hydrodynamic equations of motion, of course, must be supplemented
by specified initial  temperature and
flow velocity fields. Conventionally, these are assumed to be specified
on a thermalization proper time $\tau\sim 0.6$ fm hypersurface~\cite{Huovinen:2001cy,Heinz:2004ar}.  The agreement with
data mentioned previously is achieved using the optical Glauber model
to calculate the initial distribution of produced matter in the plane
transverse to the beam direction in conjunction with
Brodsky-Gunion-Kuhn (BGK)~\cite{Brodsky:1977de,Adil:2005qn} dynamics
that distribute initial matter in a peculiar
``trapezoidal'' manner in the longitudinal rapidity
direction~\cite{Hirano:2005xf,Wang:1991ht}.  

There exist, however, alternative and theoretically preferable models
for predicting the initial conditions of ultrarelativistic heavy ion
reactions that take into account the physics of gluon saturation at small Bjorken $x$. Color Glass Condensate (CGC) refers to a set of models that aim to
calculate QCD interactions at asymptotically large energies (at very
small Bjorken $x\ll 1$) where gluonic degrees of freedom are dominant and
the large occupation numbers of gluons provide a classical field
description of the physics; the so-called gluon saturation
regime~\cite{McLerrVenu:1994,Iancu:2000hn,BalitsKovch:9699,JIMWLK}.
Calculations of physical interest (e.g. unintegrated gluon
distribution functions, total cross sections) can be attempted
perturbatively because the characteristic scale at which non
linearities become important, $Q_{s}$, provides a large momentum scale
relative to $\Lambda_{{\rm QCD}}$.  

An implementation of the
Kharzeev-Levin-Nardi (KLN)~\cite{Kharzeev:2004if} parameterization of
the unintegrated gluon distributions from CGC is used
in~\cite{Hirano:2004rs,Hirano:2005xf} to show that CGC/KLN initial
conditions, in conjunction with ideal hydrodynamics in the QGP stage,
severely {\it overpredict} the low $p_\perp$ data at RHIC.  This is
explained by the fact that the CGC approach generically predicts a
significantly larger initial spatial eccentricity of the produced matter
 than Glauber/BGK models~\cite{Drescher:2006pi,Drescher:2006ca}.  This larger initial spatial
 eccentricity is efficiently transferred by ideal hydrodynamic motion
 into a large elliptical anisotropy in momentum space; a large $v_2$.
 Thus, if the CGC initial conditions are the ones that actually reign
 in HIC at RHIC, dissipative effects would have to be included in QGP
 evolution because hadronic dissipation has already been taken into
 account~\cite{Hirano:2005xf}.  Perturbative QCD cross sections could
 then possibly generate the amount of collectivity required at RHIC.
 
There remain, however, serious conceptual problems in the original
KLN implementation of the CGC.  As pointed out
in~\cite{Drescher:2006pi,Drescher:2006ca,Adil:2005bb,Lappi:2006xc},
the definition of $Q_s$ in the model is non-universal because it is defined
to be proportional to the Glauber participant density, which
implicitly depends on the partner nucleus.  This same problem leads to
a drop in the nuclear saturation momentum swiftly to zero as one goes
to the nuclear edge, as opposed to the correct lower limit of a single
nucleon saturation momentum~\cite{Drescher:2006pi,Drescher:2006ca}.
There are also further problems stemming from the uncertainty in the
evolution and proper treatment of the saturation momentum as one
increases Bjorken $x$~\cite{Adil:2005bb}.  The detailed properties of
the {\it local} $Q_s$ as one progresses to the nuclear edge are
crucial to the determination of the correct initial state predictions
from CGC, especially in peripheral collisions.

In this paper, we extend the {\it factorized} KLN (fKLN) approach
introduced in~\cite{Drescher:2006ca} (what we call the {\it extended local} fKLN (elfKLN)) in order to correctly calculate
the local density evolution of $Q_s$ and consistently calculate the
multiplicity of charged particles, $dN_{{\rm ch}}/d\eta$, for multiple
systems and multiple collision energies.  We take special care to account for a consistent evaluation of the model for different system types and sizes and show that this reproduces the
multiplicities in both A-A and d-A collisions at RHIC with no varying of input parameters.  The self consistency conditions force a particular COM energy dependence on the multiplicities calculated from the model which lead to surprisingly low predictions for Pb+Pb collisions at LHC energies.  These predictions are similar to ones made from simple linear extrapolations with energy of particle production from the PHOBOS collaboration~\cite{Back:2006yw}.  We do not consider the high Bjorken $x$ evolution of
$Q_s$ as it does not effect determinations of bulk multiplicity and is
better probed with more differential obervables, e.g. detailed
directed flow $v_1$ at high $p_\perp$~\cite{Adil:2005bb}.

\section{The Saturation Momentum and Unintegrated Gluon Distribution Functions}

The basic object required to calculate physical observables in the
$k_\perp$-factorized~\cite{GLR} picture employed by the gluon
saturation models is the unintegrated Gluon Distribution Function
(uGDF), $\phi_A(x,{\bf k}_\perp)$.  In principle, the uGDFs possess a
Bjorken $x$ dependence determined by nonlinear evolution equations of
the CGC theory~\cite{Iancu:2000hn,BalitsKovch:9699,JIMWLK} and their
$k_{\perp}$ dependence is fixed by a characteristic saturation
momentum, $Q_{s}(x)$.  In the McLerran-Venugopalan approach
\cite{McLerrVenu:1994} the gluon distribution is suppressed below the
saturation scale $\phi_A \sim\log({Q_{S}^{2}}/{k_{\perp}^{2}})$
compared to the perturbative form $\phi_A\sim {k_{\perp}^{-2}}$.  The
parameterization of the KLN model as used in \cite{Hirano:2004rs} is
similar to the following Lorentzian form of $\phi_{A,B}$.
\begin{equation}
\phi_{A}(x,{\bf k}_{\perp};{\bf x}_{\perp})=\frac{\kappa}{\alpha_{s}(Q_{s,A}^{2})}\frac{Q_{s,A}^{2}}{k_{\perp}^{2}+Q_{s,A}^{2}}.
\label{uGDFKLN}
\end{equation}
 The QCD coupling,
$\alpha_{s}$, is regulated at lower scales
by imposing a maximum value $\alpha_{\textrm{max}}=0.5$.  The constant
$\kappa$ is a parameter set to reproduce $dN_{ch}/d\eta$
as seen in experiment.  The transverse coordinate
dependence is implicit in the saturation momentum determined
numerically for each nucleus.

In the KLN model, the saturation scale is assumed to be proportional to the participant density in a nuclear collision.
\begin{eqnarray}
Q_{s,A}^{2}(x,{\bf x}_{\perp};{\bf b})=\frac{2\pi^{2}}{C_{F}}\alpha_{s}(Q_{s,A}^{2})xG_{\textrm{nuc}}(x,Q_{s,A}^{2})\rho_{{\rm Part},A}({\bf x}_{\perp};{\bf b}) \nonumber \\
\label{SatKLN}
\end{eqnarray} 
 In Eq.~\ref{SatKLN}, $C_{F}=\frac{N_{C}^{2}-1}{2N_{C}}$ %is the SU($\mathrm{N_{C}}$) Casimir 
and $x$ is the collinear momentum fraction given by kinematics.  The participant density is defined in terms of the usual Glauber thickness functions $T_{A/B}({\bf x}_\perp)$ as in~\cite{Hirano:2004rs,Kharzeev:2004if}.  The {\it integrated} Gluon Distribution Function (iGDF) is calculated to be (in the perturbative regime),
\begin{equation}
xG_{\textrm{nuc}}(x,Q^{2})=K\log(\frac{Q^{2}+\Lambda^{2}}{\Lambda^{2}})x^{-\lambda}(1-x)^{n}
\label{xG}
\end{equation}
In Eq.~\ref{xG}, the $x^{-\lambda}$ term accounts for the rapid growth
of small Bjorken $x$ gluons while the factor of
$(1-x)^{n}$ was introduced in KLN to account qualitatively
for the rapid depletion of gluons as
$x\rightarrow 1$ outside the small $x$ framework
of the CGC model.  The unintegrated gluon distribution shown in~\ref{uGDFKLN} is defined such that $\phi(x,{\bf k}_\perp)=\frac{\partial(xG)}{\partial\log k_\perp^2}=k_\perp^2\frac{\partial(xG)}{\partial k_\perp^2}\sim Q_s^2\frac{\partial(xG)}{\partial k_\perp^2}$.  This relationship holds in the leading logarithm DGLAP kinematic regime where $k_\perp^2\gg Q_s^2$.  The KLN model has the freedom to set the normalization of $xG(x,Q^2)$ independently of $\phi(x,{\bf k}_\perp^2)$ due to the $K$ factor in Eq.~\ref{xG} in the leading log approximation.  This variable is set so as to reproduce the experimentally implied value of $Q_s$.  We will specify the numerical value of all the parameters after discussing the adjusted KLN type saturation model that we use in the paper.  Our model differs from both the KLN and fKLN implementations of the CGC in this case, as we self consistently include the high $x$ depletion factor of $(1-x)^n$ in both $\phi_A(x)$ and the determination of $Q_s$.  Other models only include this factor in the uGDFs, leading to a relative depletion of particle production in the high $x$ regions in our model relative to previous ones.  This feature will be essential to our description of asymmetric nuclear collisions later in this paper.

\begin{figure}[t!]
%\hspace*{-.6in}
\epsfig{file=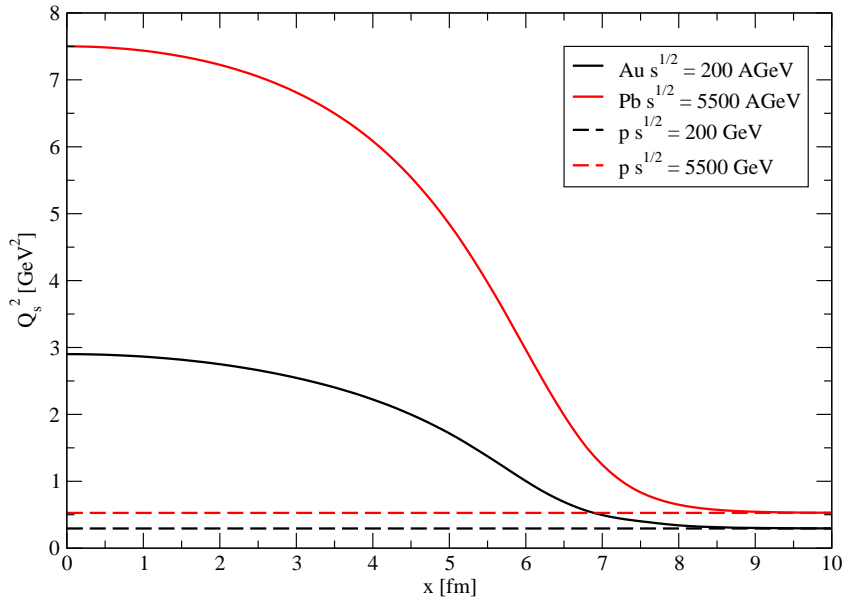, width=4in,angle=270}
\caption{(Color Online) The figure shows the squared saturation momentum as a function of transverse coordinate for both an Au nucleus at Bjorken $x=0.01$ and a Pb nucleus at Bjorken $x=0.01\frac{200}{5500}\sim10^{-4}$.  The horizontal lines are the equivalent nucleon saturation momenta shown to guide the eye.
}
\label{QSat}
\end{figure}
As can be seen in Eq.~\ref{SatKLN}, the saturation momentum defined in the KLN model is not universal as it explicitly depends on the partner nucleus through its dependence on $\rho_{{\rm Part}}({\bf x}_\perp)$~\cite{Drescher:2006ca,Lappi:2006xc,Adil:2005bb}.  The problem at hand is to modify the constitutive equations of the KLN model such that the saturation momentum depends only on the Glauber thickness function $T_{A/B}({\bf x}_\perp)$ of the nucleus in question and still reproduces the phenomenologically successful participant dependence of the multiplicity that follows naturally from the conventional KLN model~\cite{Kharzeev:2004if}.  

Just such an adjustment is detailed in~\cite{Drescher:2006ca} where they create the fKLN model.  They show that the correct event averaging of the uGDF involved in a nuclear collision leads to the replacements $Q_s^2(\rho_{{\rm Part}})\rightarrow Q_s^2(T_A/p_A)$ and $\phi\rightarrow p_A\phi$ (using the notation found in~\cite{Drescher:2006ca}) where $p_A({\bf x}_\perp)\equiv (1-(1-\sigma_{NN}T_A({\bf x}_\perp)/A)^A)$ is the probability of finding at least one excited nucleon at the position ${\bf x}_\perp$ in nucleus $A$.  The inelastic nucleon-nucleon cross section is set to $\sigma_{NN}(\sqrt{s}=200,5500)=42,70$ mb.  Explicitly writing out the changes,
\begin{eqnarray}
Q_{s,A}^{2}(x,{\bf x}_{\perp})=\frac{2\pi^{2}}{C_{F}}\alpha_{s}(Q_{s,A}^{2})xG_{\textrm{nuc}}(x,Q_{s,A}^{2})\frac{T_{A}({\bf x}_{\perp})}{p_A({\bf x}_{\perp})}, \nonumber \\
\phi_{A}(x,{\bf k}_{\perp};{\bf x}_{\perp})=p_A({\bf x}_{\perp})\frac{\kappa}{\alpha_{s}(Q_{s,A}^{2})}\frac{Q_{s,A}^{2}}{k_{\perp}^{2}+Q_{s,A}^{2}}.
\label{fKLN}
\end{eqnarray}
Note that we no longer have the problems occurring from the $Q_s\rightarrow0$ limit that were found in~\cite{Adil:2005bb} because the $A\rightarrow1$, or alternatively the $T_A\rightarrow0$, limit leads to a finite saturation momentum equal to that of a single nucleon.
\begin{equation}
Q_{s,p}^{2}(x)=\frac{2\pi^{2}}{C_{F}}\alpha_{s}(Q_{s,p}^{2})xG_{\textrm{nuc}}(x,Q_{s,p}^{2})\frac{1}{\sigma_{NN}}
\label{Qsp}
\end{equation}
We use a value of $\lambda=0.288$ as in~\cite{Kharzeev:2004if} and set the value of the constant $K=0.56$ in Eq.~\ref{xG} to reproduce $\langle Q_{s,A}^2(x=0.01)\rangle\sim2$ GeV$^2$ as in ~\cite{Hirano:2004rs,Kharzeev:2004if}.  The averaging is over the transverse coordinate plane using $T_{A}$ as weight.

Fig.~\ref{QSat} shows the squared saturation momentum as a function of the transverse coordinate at given value of Bjorken scaling variable, $x$.  One can easily see that as one nears the nuclear edge, the nuclear saturation momentum tends towards the nucleon saturation momentum from above.  An added nice feature of the fKLN approach is that the nucleon saturation momentum is set uniquely by the low density limit of the nucleus saturation momentum and is not a free parameter as in~\cite{Kharzeev:2004if}.

\section{Multiplicity in Symmetric Collisions} 

We first concentrate on nucleon-nucleon collision and can calculate the multiplicity of charged particles in p-p$(\bar{{\rm p}})$ collisions using the formula of Gribov-Levin-Ryskin~\cite{GLR}.
\begin{equation}
  \frac{dN_{ch/pp}}{d^2{\bf p}_{\perp}dy}=\frac{2}{C_{F}}\frac{\alpha_{s}(p_\perp^{2})}{p_{\perp}^2}
\int^{p_\perp} d^{2}{\bf k}_{\perp}  \phi_{p}(({\bf k}_{\perp}+{\bf p}_{\perp})/2;Q_{s,p}^2(x_1)) \phi_{p}(({\bf k}_{\perp}-{\bf p}_{\perp})/2;Q_{s,p}^2(x_2)).
 \label{eqn:ktfac}
\end{equation}
In Eq.~\ref{eqn:ktfac}, the collinear momentum fractions are given by kinematics as
$x_{1,2}=p_{\perp}\exp(\pm y)/\sqrt{s}$.  Eq.~\ref{eqn:ktfac} can be integrated over ${\bf p}_\perp$ to get the multiplicity of charged particles in p-p collisions, $dN_{ch/pp}/dy$.

\begin{figure}[t!]
%\hspace*{-.6in}
\epsfig{file=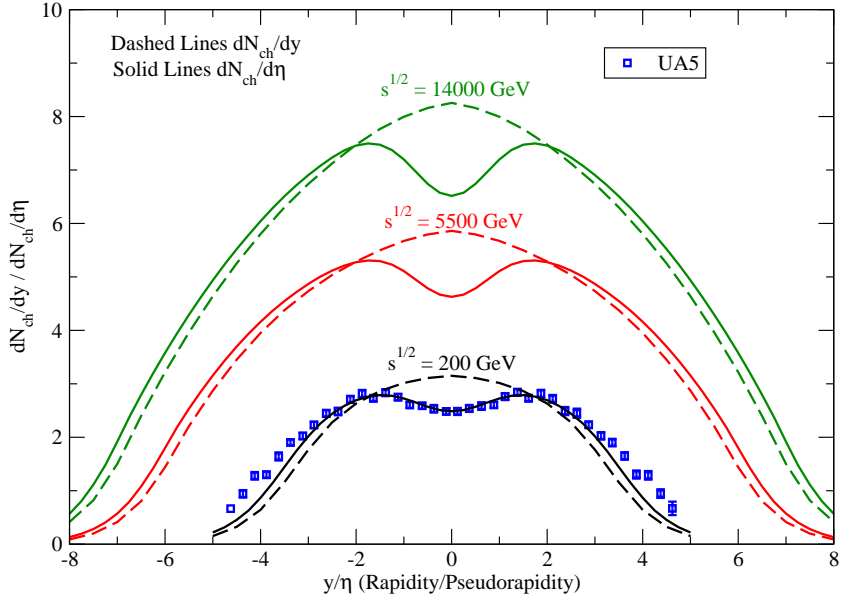, width=4in,angle=270}
\caption{(Color Online)  The rapidity and pseudorapidity density of produced charged particles in nucleon-nucleon collisions are shown as a function of the rapidity/pseudorapidity.  The model is tuned to reproduce data from the UA5 collaboration~\cite{Alner:1986xu} and then predictions are made for the charged multiplicity at LHC energies of $\sqrt{s}=5500,14000$ GeV.
}
\label{ppmult}
\end{figure}
The value of $\kappa\sim0.02$ is set to reproduce data from the UA5 collaboration~\cite{Alner:1986xu}.  One subtlety in the analysis is the role played by the transformation from the rapidity $y$ to the pseudorapidity $\eta$.  The conventional Jacobian is used as in~\cite{Kharzeev:2004if} but there is the matter of setting the parameter $m_g/p_\perp$ found in the Jacobian.  We use the experimentally determined average transverse momentum in the collision as a measure for $p_\perp$~\cite{Adams:2004cb} and find that the best fit is found when $m_g\sim0.35$ GeV.  We use the experimentally found average $p_\perp$~\cite{Adams:2004cb} in the Jacobian for all collisions at $\sqrt{s}=200$ GeV and assume that the ratio $m_g/p_\perp$ does not change as one goes to higher energies.  The calculation for $dN_{ch}/dy$ is shown alongside $dN_{ch}/d\eta$ in the figured to show the effect of the Jacobian.

Fig.~\ref{ppmult} shows the results of evaluating $dN_{ch/pp}/dy$ for different COM energies.  One can see that the multiplicity is described well at $\sqrt{s}=200$ GeV except at very high rapidities.  This will be a recurring feature of the calculation as large rapidites are in the fragmentation region of the collision and are not well modeled by gluon saturation physics.  The theory has been fit to the UA5 data~\cite{Alner:1986xu} and all other calculations are predictions as there are no more parameters to tune.  The predictions for the p-p collisions at LHC energies can also be seen in Fig.~\ref{ppmult}.

\begin{figure}
\centering
 \epsfig{file=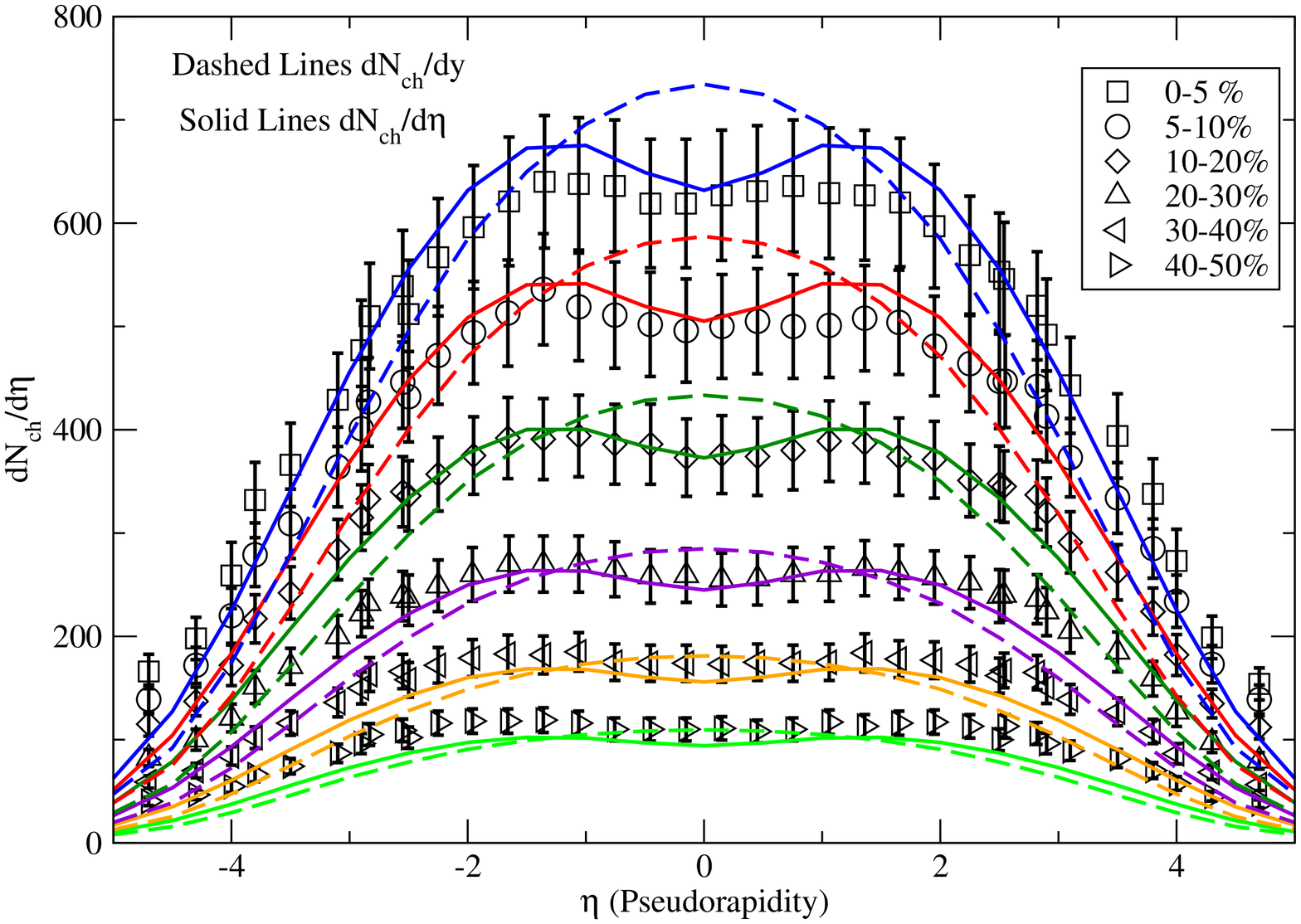,width=2.5in,angle=270}
 \epsfig{file=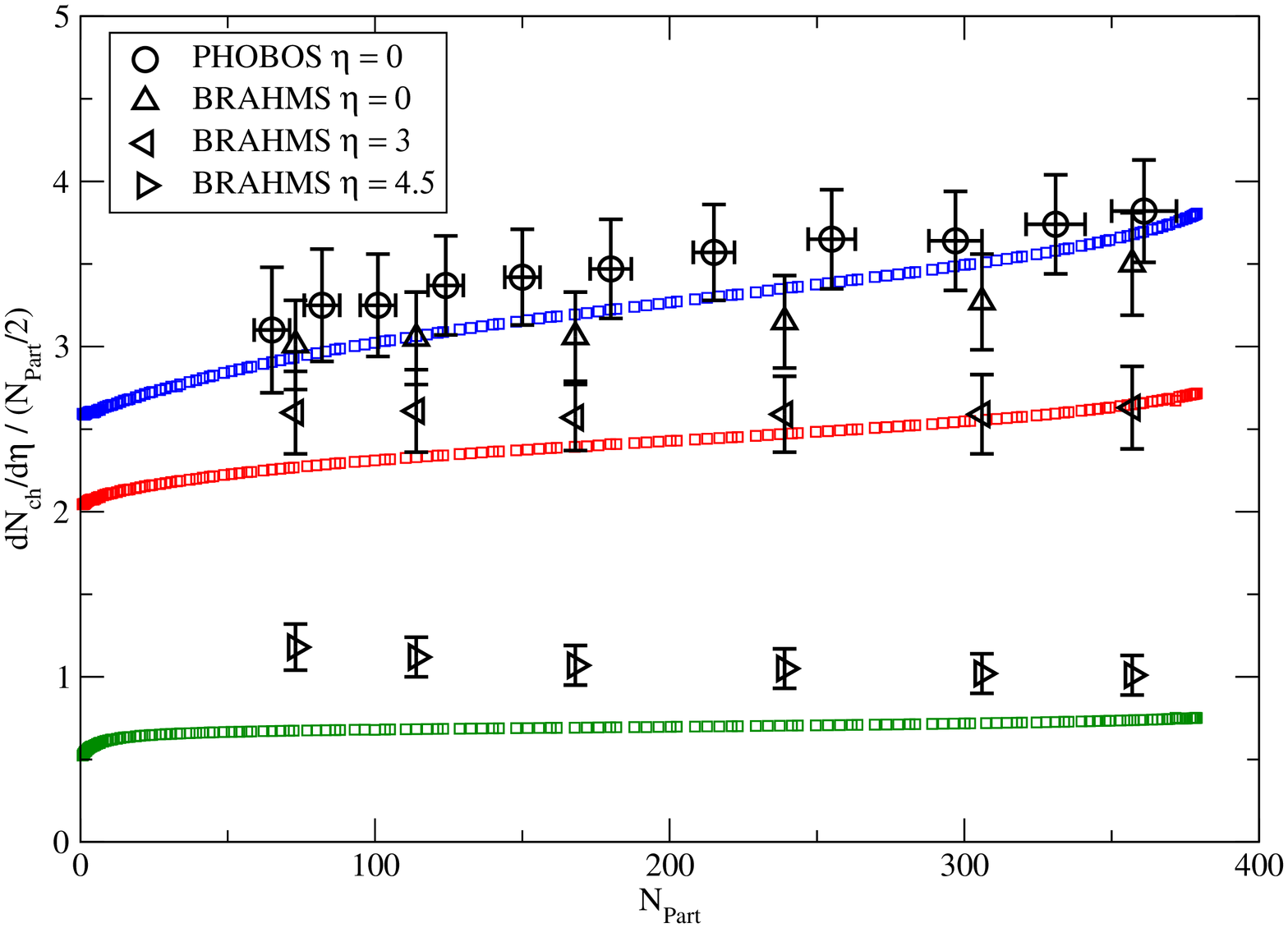,width=2.5in,angle=270}
  \caption{(Color Online) Left : Charged multiplicity distribution as a function of rapidity/pseudorapidity for different centrality bins in $\sqrt{s}=200$ GeV Au-Au collisions at RHIC.  Data includes both statistical and systematic errors and is from the BRAHMS Collaboration~\cite{Bearden:2001qq}.  Right:  The figure shows charged particle production per participant pair as a function of the number of participants in the collision.  Data from BRAHMS~\cite{Bearden:2001qq} and PHOBOS~\cite{Back:2002uc}.
}
  \label{AuAuRHIC}
\end{figure}

\begin{figure}
\centering
 \epsfig{file=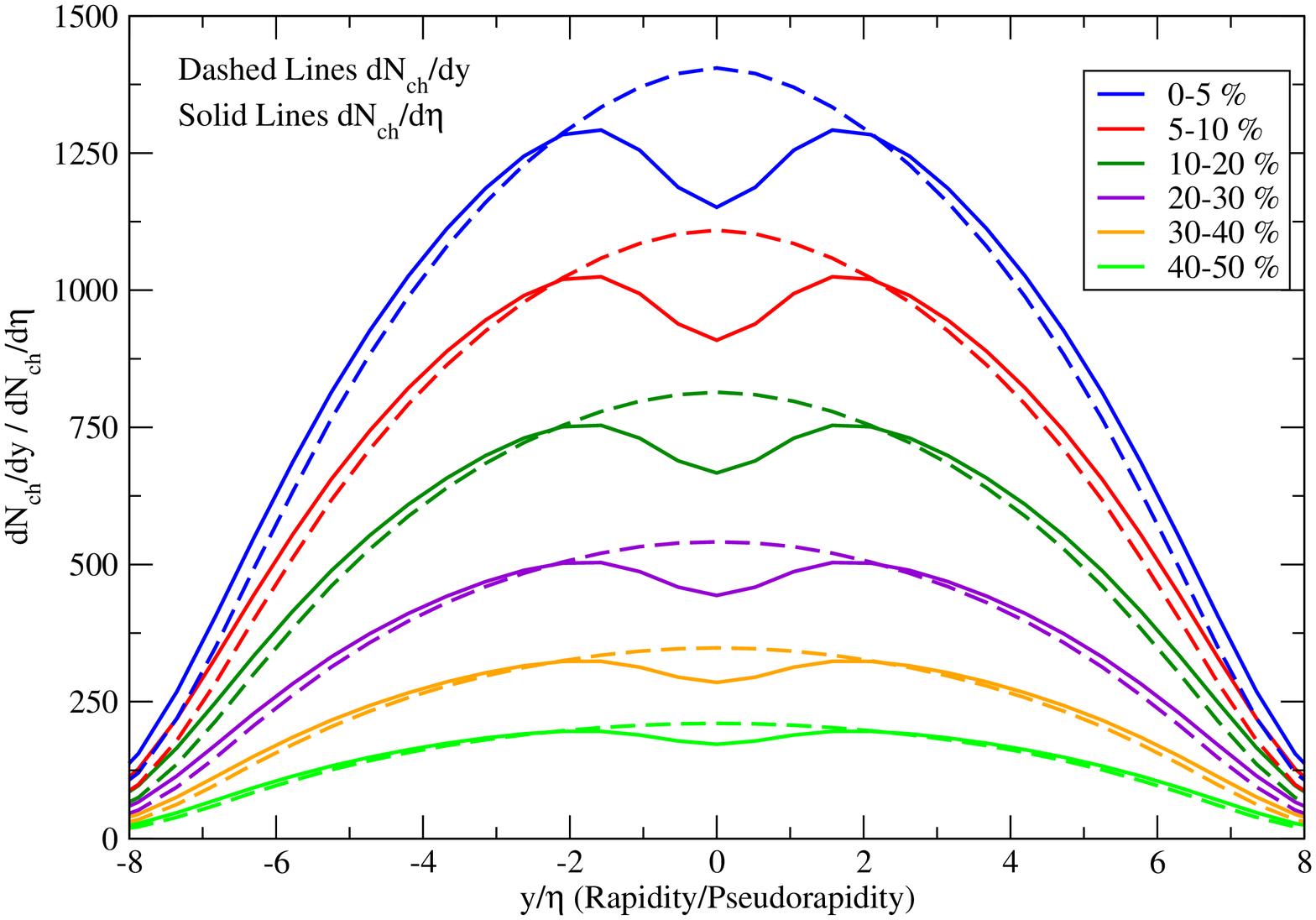,width=2.5in,angle=270}
\epsfig{file=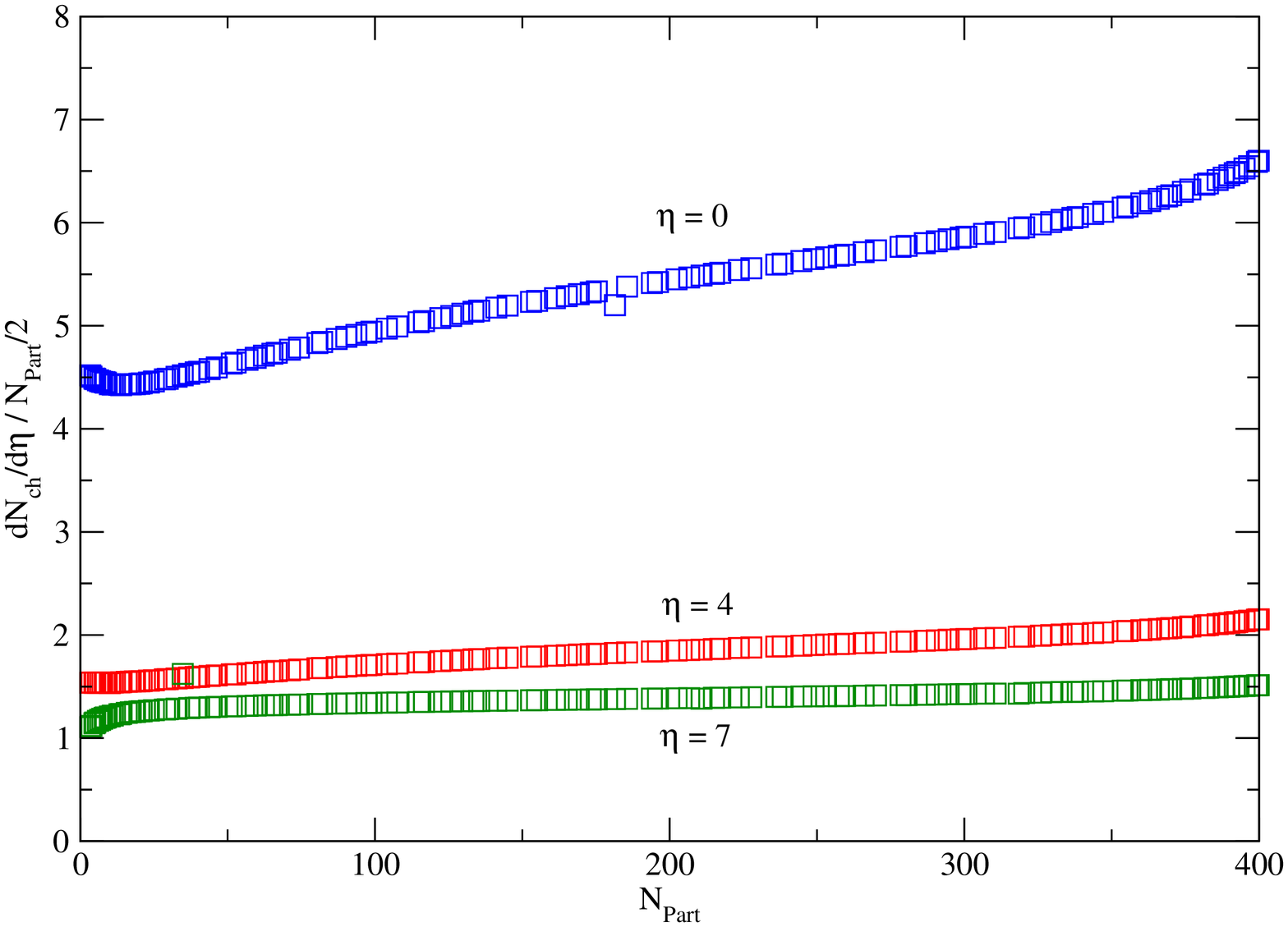,width=2.5in,angle=270}
  \caption{(Color Online) Left : Charged multiplicity distribution as a function of rapidity/pseudorapidity for different centrality bins in $\sqrt{s}=5500$ GeV Pb-Pb collisions at LHC.  Right:  The figure shows charged particle production per participant pair as a function of the number of participants in the collision at LHC. }
    \label{PbPbLHC}
\end{figure}

Once the parameters of the theory have been specified by reproducing p-p data, we move on to consider symmetric nuclear collisions;  Au-Au collisions at the max RHIC energy of $\sqrt{s}=200$ AGeV and Pb-Pb collisions at the expected LHC energy of $\sqrt{s}=5500$ AGeV.    When we move over to a nuclear collision, we must treat particle production locally as a density in the plane transverse to the beam.  The parameter values, in particular the normalization of the uGDF, is set to produce the correct number of particles in a p-p collision.  In order to interpret this as a local transverse plane density, we note that the particle production in Eq.~\ref{eqn:ktfac} is implicitly spread out over a transverse area the size of the nucleon-nucleon cross section, $\sigma_{NN}$.  We can further understand the smearing of particle production over an area of $\sigma_{NN}$ by noting (as shown in Eqs.~\ref{SatKLN} and ~\ref{fKLN}) that all nucleons present in an interaction region of size $\sigma_{NN}$ are considered part of a {\it local} density that determines the $Q_s$ for particle production.  We convert Eq.~\ref{eqn:ktfac} into a local transverse space density,
\begin{equation}
\frac{dN_{ch,pp}}{d^2{\bf x}_\perp d^2{\bf p}_\perp dy}=\frac{1}{\sigma_{NN}}\frac{dN_{ch,pp}}{d^2{\bf p}_\perp dy}.
\label{localpp}
\end{equation}
The generalization of this to a full nuclear A+B collision is immediate.
\begin{equation}
  \frac{dN_{ch/AB}}{dy}(\sqrt{s})=\frac{1}{\sigma_{NN}(\sqrt{s})}\int d^2{\bf x}_\perp\int d^2{\bf p}_{\perp}\frac{2}{C_{F}}\frac{\alpha_{s}(p_\perp^{2})}{p_{\perp}^2}
\int^{p_\perp} d^{2}{\bf k}_{\perp}  \phi_{A}(({\bf k}_{\perp}+{\bf p}_{\perp})/2;Q_{s,A}^2(x_1)) \phi_{B}(({\bf k}_{\perp}-{\bf p}_{\perp})/2;Q_{s/B}^2(x_2)).
\label{ktfacAB}
\end{equation}
We regard Eq.~\ref{ktfacAB} as the defining expression for our {\it extended local} factorized KLN (elfKLN) model of the CGC.  It includes the consistent local $A\rightarrow1$ limit implemented by fKLN in addition to the self consistent high $x$ depletion factor of $(1-x)^n$ included in both the uGDFs and $Q_s$ determination.  This change makes it possible to set a universal normalization of the uGDFs such that the adjustments in going from $\phi_p$ to $\phi_A$ are twofold; we replace $Q_{s,p}$ with $Q_{s,A}$ and introduce a factor of $p_A$ in the nuclear distribution.

The left panel in Fig.~\ref{AuAuRHIC} shows the predictions for $dN_{ch}/dy$ and $dN_{ch}/d\eta$ in Au-Au collisions at maximum RHIC energies along with data from the BRAHMS Collaboration~\cite{Bearden:2001qq} as a comparison.  One can see that we achieve a very good description of the data over all centrality classes with the exception of the regions near the projectile and target rapidity where fragmentation effects dominate.  The centrality cuts are imposed via a geometric Glauber analysis as in~\cite{Kharzeev:2004if}.  We do not take into account the effects of the event by event fluctuation in geometry caused by considering the Monte Carlo Glauber analysis which has been shown to be important in explaining system size dependence of observables for smaller systems (especially eccentricity and elliptic flow with fluctuations)~\cite{Alver:2007rm,Sorensen:2006nw,Alver:2006wh}.

The right panel in Fig.~\ref{AuAuRHIC} shows more clearly the system size dependence of particle productions at RHIC.  It shows the charged particle production per participant pair as a function of $N_{{\rm Part}}$ for a number of different pseudorapidities.  We get a good description of the particle production, as seen by the agreement between our calculations and data from BRAHMS~\cite{Bearden:2001qq} and PHOBOS~\cite{Back:2002uc}.  Note that, unlike previous attempts at reaching the $N_{{\rm Part}}/2\rightarrow1$ limit~\cite{Kharzeev:2004if}, there is no problem in this calculation due to the consistent treatment of the local $Q_s$.  The calculation extrapolates right to the charged multiplicity in a single p-p collision as one reduces $N_{{\rm Part}}$.  This agreement is non trivial and is a good check of our approach as all parameters have already been fit and self consistency demands that we get the correct limit.

The left panel of Fig.~\ref{PbPbLHC} shows the rapidity and pseudorapidity densities calculated from the integral of Eq.~\ref{ktfacAB} in the case of a Pb-Pb collision at $\sqrt{s}=5500$ AGeV.  The centrality cuts are performed using optical Glauber models as in~\cite{Kharzeev:2004if}.  The predictions for LHC multiplicity are surprisingly lower than some previous predictions from saturation physics~\cite{Kharzeev:2004if}, even the ones using an fKLN type model~\cite{Drescher:2007ax}.  The cause of this difference can be tracked to the form for Eq.~\ref{ktfacAB} imposed upon us by the self consistency we require between p-p and A-A collisions.  Since we have a purely {\it local} description of particle production (which ~\cite{Kharzeev:2004if} does not include), we are forced to include the factor of $1/\sigma_{NN}$ to maintain consistency between Eqs.~\ref{eqn:ktfac} and~\ref{ktfacAB}.  Therefore, we are sensitive to two processes that increase total particle production as one increases COM energy; 1) the increase of gluon production {\it per unit area} caused by small $x$ evolution, and 2) the increase of the underlying interaction cross section $\sigma_{NN}(\sqrt{s})$ used as an ``averaging area'' for coherent interaction (as can be seen by the presence of $\sigma_{NN}$ in the definition of the local density $T_A({\bf x}_\perp)$).  Previous calculations include 1) but not 2) and are thus typically larger by a factor of 70/42 $\sim$ 1.66 than the current predictions.  

\begin{figure}
\centering
 \epsfig{file=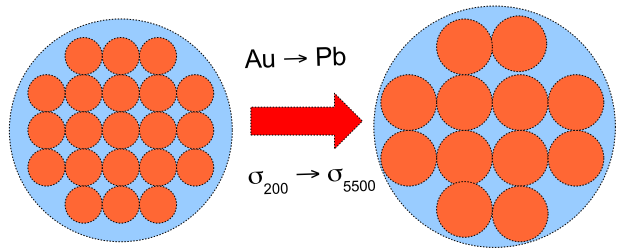,width=3in,angle=90}
  \caption{(Color Online) A schematic figure showing the changes in the local geometric averaging as one goes from Au+Au collisions at RHIC energies to Pb+Pb collisions at LHC energies.  Both the increase in size of the nucleus and the typical coherence area are shown. }
    \label{AutoPb}
\end{figure}
The reason for this difference is schematically shown in Fig.~\ref{AutoPb}.  As one increases $\sqrt{s}$, one naturally lowers the typical $x$ value being probed in the interaction and thus increases the typical saturation scale.  This leads to more particle production at higher energies as compared to lower energies.  The coherence effect, however, is implicitly spread out over an area the size of $\sigma_{NN}(\sqrt{s})$.  As schematically shown in Fig.~\ref{AutoPb}, this increase in the coherence region limits the number of coherent production points one can ``fit'' into the interactions region, thus leading to a suppression in overall particle production.  

Our elfKLN model consistently includes both effects mentioned.  We systematically isolate the production per unit area by using Eq.~\ref{ktfacAB} and thus can consistently explain both p-p and A-A data at RHIC energies, which previous approaches are unable to do without invoking an arbitrary enhancement for their p-p predictions~\cite{Kharzeev:2004if,Drescher:2007ax}.  The tendency is similar to the one found at RHIC and we note the consistent p-p system limit as we saw in the RHIC calculations.  The right panel shows the particle production per participant pair at LHC energies as a function of $N_{{\rm Part}}$.

\section{Multiplicity in Asymmetric Collisions}

\begin{figure}
 \epsfig{file=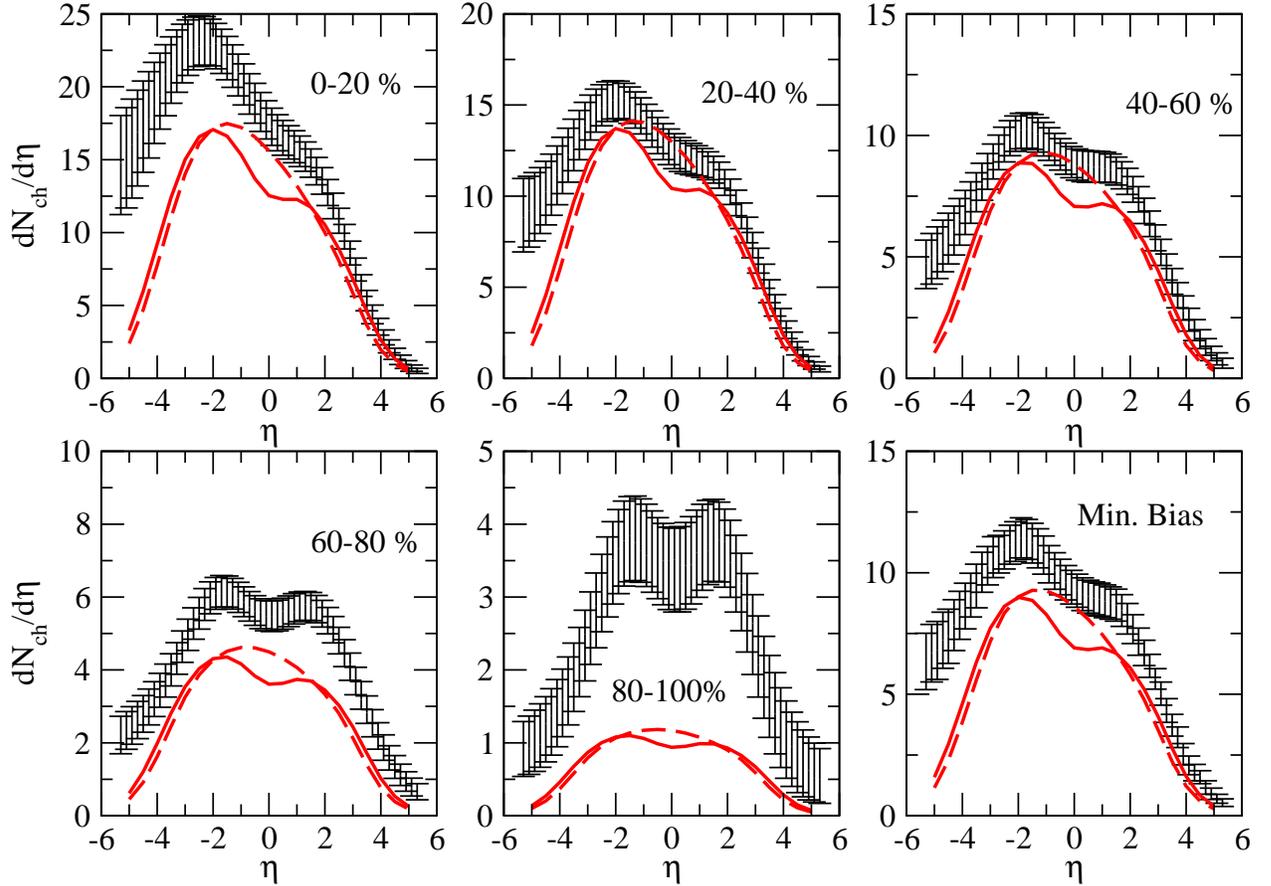,width=6in,angle=270}
  \caption{(Color Online) The Figure shows a comparison of the multiplicity as a function of the rapidity/pseudorapidity for different centralities as measured by the PHOBOS Collaboration~\cite{Back:2004mr} and calculated in this paper.  Dashed lines signify $dN_{ch}/dy$ while solid lines are $dN_{ch}/d\eta$.}
    \label{dAuAbs}
\end{figure}

Eq.~\ref{ktfacAB} can be directly generalized to the case of asymmetric collisions by inserting the Glauber thickness functions of the different collision partners used.  For the particular case of deuteron-Au collisions at maximum RHIC energies, one needs to use a model of the nuclear density associated with the deuteron nucleus.  We use the Hulthen distribution to describe the deuteron, using both $s-$ and $d-$ wave contributions as in~\cite{Kharzeev:2004if}.  Once the distribution is determined, we calculate the Glauber thickness function $T_A({\bf x}_\perp)$ and proceed as before.

Fig.~\ref{dAuAbs} shows the comparison of our calculation with data from the PHOBOS Collaboration~\cite{Back:2004mr} for a number of different centrality cuts as well as for the minimum bias data.  The first thing to notice is that there is very little agreement in the total number of particles produced at any centrality.  The theoretical calculation consistently under predicts the data, so much as predicting a multiplicity for a centrality of $80-100 \%$ that is below the multiplicity for a p-p collision.  An interesting point to note, however, is that the shape of the overall data is very well reproduced just from the asymmetry of saturation momenta in the collision.  The presence of the factor of $(1-x)^n$ in the determination of $Q_s$ is essential to the reproduction of this asymmetry as it causes a precipitous drop in $Q_s$ at high rapidities.  Previous models~\cite{Kharzeev:2004if} achieve asymmetric results by supplementing their far too slowly varying $Q_s$ determinations with a leading logarithm evaluation for the $p_\perp$ integral in Eq.~\ref{ktfacAB}.  This picks the points in phase space that possess the most asymmetric values of $Q_s$ between the two collision partners.
\begin{center}
\begin{tabular}{|c|c|c|c|c|}
\hline
Centrality &$\langle N_{{\rm Part}}^{Au}\rangle_{op}$&$\langle N_{{\rm Part}}^{d}\rangle_{op}$&$\langle N_{{\rm Part}}^{Au}\rangle_{ex}$&$\langle N_{{\rm Part}}^{d}\rangle_{ex}$\\
 \hline
 $0-20 \%$& 11.72 &1.99 & $13.5\pm1.0$&$2.0\pm0.1$\\
 \hline
 $20-40 \%$& 8.21 &1.91 & $8.9\pm0.7$&$1.9\pm0.1$\\
 \hline
 $40-60 \%$& 4.59 &1.63 & $5.4\pm0.6$&$1.7\pm0.2$\\
 \hline
 $60-80 \%$& 1.96 &1.00 & $2.9\pm0.5$&$1.4\pm0.2$\\
 \hline
 $80-100 \%$& 0.47 &0.29 & $1.6\pm0.4$&$1.1\pm0.2$\\
 \hline
 minbias & 5.37 &1.37 & $6.6\pm0.5$&$1.7\pm0.1$\\
 \hline
\end{tabular}
\end{center}
We note that the determination of centrality cuts in our calculations was performed using an optical Glauber analysis as in~\cite{Kharzeev:2004if}.  As previously shown in~\cite{Kharzeev:2004if,Alver:2006wh} the deviations between a mean field treatment as in the current paper and a fully fluctuating Monte Carlo treatment as performed by the experiment is significant, specially when one is studying smaller systems.  One way to determine the difference between the two approaches is to tabulate the average number of participants calculated in each, as shown in the table above.  The experimental data in the table is taken from the PHOBOS Collaboration~\cite{Back:2004mr}.  One can see that there is an appreciable difference in the participant number between the two approaches, with the fluctuating approach isolating larger systems (i.e. smaller impact parameters) for each centrality cut.  The table also explains the peculiarity found in Fig~\ref{dAuAbs} for the $80-100\%$ centrality cut.  The optical Glauber actually gives total participant number of less than 1 for that centrality cut, thus leading to particle production below the p-p level.

\begin{figure}
 \epsfig{file=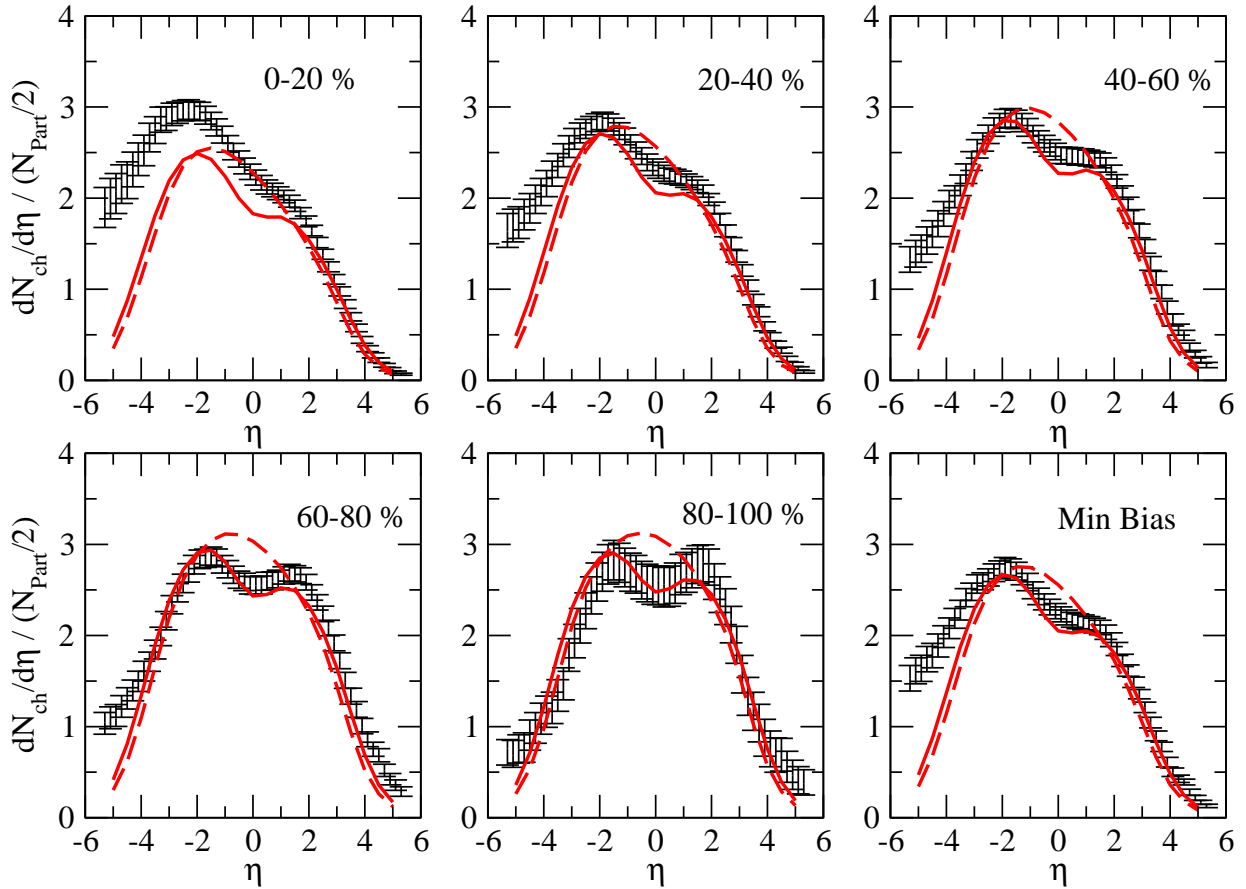,width=6in,angle=270}
  \caption{(Color Online) The Figure shows a comparison of the multiplicity per participant pair as a function of the rapidity/pseudorapidity for different centralities as measured by the PHOBOS Collaboration~\cite{Back:2004mr} and calculated in this paper.  Dashed lines signify $dN_{ch}/dy$ while solid lines are $dN_{ch}/d\eta$.}
\label{dAuRat}
\end{figure}
Given the different determinations of centrality in the approaches, a good way to compare the data to the calculation is to divide out by the overall number of participant pairs.  This controls for the difference in the underlying impact parameter differences and poses the problem in terms of particle production per participant pair.  The analysis is presented in Fig.~\ref{dAuRat} and it can be seen that the comparison is a lot more favorable now that differences due to centrality determination have been removed.  There is good agreement between data and theory, specially in the mid to forward rapidity region.  There is still some discrepancy as one moves backward in rapidity towards the nucleus.  Intranuclear cascading increases the multiplicity of charged particle in the nucleus fragmentation region.  Furthermore, we have neglected to take into account the geometric scaling region of the uGDF at low $x$~\cite{Levin:2000mv} which has been shown to be important in d-Au collisions at RHIC, at least off mid rapidity~\cite{Dumitru:2005gt}.  Both these effects are larger for more central collisions, a fact that is borne out in Fig.~\ref{dAuRat} where the discrepancy between data and theory is reduced as one moves towards more peripheral collisions.

\begin{figure}
\centering
 \epsfig{file=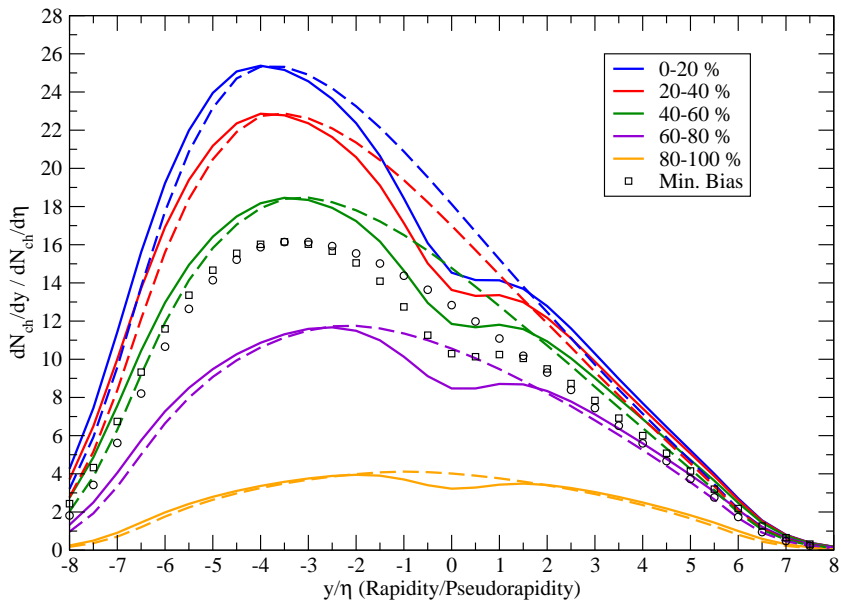,width=4in,angle=270}
  \caption{(Color Online) The figure shows the rapidity distribution of charged particles produced in a p-Lb collision at $\sqrt{s}=5500$ GeV.  Dashed lines denote $dN_{ch}/dy$ while solid lines are $dN_{ch}/d\eta$.  The open squares are minimum bias predictions.}
    \label{pPb}
\end{figure}
Having achieved a good description of d-Au data at RHIC we move on to a prediction for p-Pb collisions at LHC energies.  Eq.~\ref{eqn:ktfac} easily generalizes to an asymmetric p-Pb collisions by replacing one of the uGDFs with one that uses the saturation momentum of a nucleus.  Fig.~\ref{pPb} shows predictions of a p-Pb collision for different centralities at $\sqrt{s}=5500$ GeV.  The overall charged particle production is smaller by a factor of about 2 from p-Lb predictions made using the conventional KLN model~\cite{Kharzeev:2004if}.  One must note, however, that this calculation makes centrality cuts using an optical Glauber model calculation and thus will run into similar problems to the d-Au calculations when being compared to data.  The problems can be seen explicitly in the the $80-100\%$ centrality bin in Fig.~\ref{pPb}, where the particle production is at a level lower than p-p at $\sqrt{s}=5500$ GeV (Fig.~\ref{ppmult}).  The minimum bias prediction is most robust, as was the case with the d-Au predictions.  We present this prediction as a good estimate of the expected multiplicity, one that will be better compared to data when presented as particle production per participant pair.  In general, a complete description of asymmetric nuclear collisions may best be attempted using a model that can better accommodate fully fluctuating geometries, such as the Monte Carlo KLN (MCKLN) model detailed in~\cite{Drescher:2006ca}.

\section{Conclusions}

The recent claims supporting the sQGP paradigm at RHIC are heavily dependent on the success of models using ideal hydrodynamics to model the evolution of the QGP stage of bulk evolution at RHIC~\cite{Huovinen:2001cy,Heinz:2004ar}.  This success is inherently linked to the choice of initial conditions assumed in the models, and an initial state derived from gluon saturation / CGC models does not work well in conjunction with ideal hydrodynamics to explain RHIC data~\cite{Hirano:2005xf}.  This is due to the larger initial spatial eccentricity produced in gluon saturation models which is naturally converted into a larger elliptic flow parameter, $v_2$~\cite{Drescher:2006pi}.  A better understanding of the origin and robustness of this large eccentricity is needed.  In order to do this we need to understand the $A\rightarrow1$ and $T_A\rightarrow0$ limits of the theory, which is equivalent to creating a model that works in describing not just central A-A collisions but also p-p and p-A collisions in one consistent model.  This is especially important as recent work indicates that models that have more realistic nuclear edge characteristics lead to a spatial eccentricity that lies somewhere between the pure Glauber/BGK and KLN model eccentricities~\cite{Drescher:2006ca,Lappi:2006xc}.

We use the elfKLN model, our extension of the fKLN model in~\cite{Drescher:2006ca}, consistently applied to p-p, p-A and d-A collisions along with the usual description of A-A collisions.  We find that the model does an excellent job consistently describing particle production in symmetric nuclear collisions from p-p all the way up to A-A collisions.  We fix our parameters to reproduce particle production in p-p collisions and predict the multiplicity in Au-Au collisions at RHIC, finding good agreement with the data.  The model does very well in treating the nuclear edges consistently, as evidence by Fig.~\ref{AuAuRHIC} where we show a consistency check between our Au-Au and p-p calculations.  We also predict the multiplicity for Pb-Pb collisions at LHC energies, which are found to be surprisingly small.  This is due to the increasing ``coherence area'' of the interaction, which we consistently build into our model.  

We move on to calculate asymmetric nuclear collisions (d-A, p-A) in an effort to correctly model the edges of peripheral A-A collisions.  We find that absolute agreement with the data is difficult to attain if one uses optical mean field methods to obtain the centrality cuts.  However, good agreement is achieved with the d-Au RHIC data if one accounts for the variation in the underlying geometry by comparing particle production per participant pair.

In closing, we propose the elfKLN model as a robust model for bulk CGC physics that has the properties of being explicitly factorized and well defined in the diffuse nucleus limit.  We do, however, recognize that it it might not be applicable in highly asymmetric (and thus smaller system size) nuclear collisions, where a fully fluctuating model such as the MCKLN of~\cite{Drescher:2006ca} might be more appropriate.  Our prediction does not include effects such as running coupling and pre asymptotic terms in the evolution of the saturation momentum with energy~\cite{Kharzeev:2007zt,Albacete:2007hg}.  These adjustments are left for future work.

\begin{acknowledgments}
Discussions with A.~Dumitru, H.-J.~Drescher, B.~Cole, D.~Molnar, A.~Blaer, D.~Winter and J.~Jia
are gratefully acknowledged.
MG acknowledges support from
the DFG, %eutsche Forschungsgemeinschaft 
% as a Mercator Gastprofessur at 
%Frankfurt Institute for Advanced Studies 
%the 
FIAS and ITP %Institut fuer Theoretische
%Physik of
of the J.W. Goethe Uni. Frankfurt, and from %the
GSI. %Gesellschaft fuer Schwerionenforschung GSI. 
  This work is also supported in part by the United States
Department of Energy
under Grants   No. DE-FG02-93ER40764.
\end{acknowledgments}

\end{document}